\documentclass[epj,final]{svjour}
\usepackage{amssymb,amsmath,graphicx,bbm,bm,subfigure}
\begin{document}
\title{Bell violation versus geometric measure of quantum discord and their dynamical behavior }
\author{Yao Yao\and Hong-Wei Li\and Mo Li\and Zhen-Qiang Yin\and Wei Chen\and Zheng-Fu Han
}                     
%
%
\mail{zfhan@ustc.edu.cn}
\institute{Key Laboratory of Quantum Information,University of Science and
Technology of China,Hefei 230026,China
}

\date{Received: date / Revised version: date}
%
\abstract{ Motivated by recent numerous works on the interplay among various measures of quantum correlations,
we aim to investigate the relationship between the violation of Clauser-Horne-Shimony-Holt (CHSH) Bell inequality
and geometric measure of quantum discord for two-qubit systems. Exact lower and upper bounds of Bell violation
versus geometric discord are obtained for a specific and significant class of states, Bell diagonal states,
and the respective states which suffice those bounds are also characterized. The dynamical behavior of these
two quantifiers is carefully analyzed in the presence of decoherence, including Markovian, non-Markovian, and
non-back-action quantum environments. The results suggest that Bell violation is closely related to geometric discord,
like its relationship with other entanglement monotones.
} 
\maketitle

\section{Introduction}
Quantum nonlocality, revealed by the violation of Bell-type inequalities \cite{Bell64}, has already been recognized as the
fundamental resource of present-day quantum information science, and quite recently, Bell violations (especially
the violation of CHSH inequality \cite{CHSH69}) find expanding applications in new quantum information tasks, such as
device-independent random number generation (DIRNG) \cite{Pironio10} and quantum key distribution (DIQKD) \cite{Barrett05,Acin07,Pironio09}. The general relations between the violation of CHSH inequality and entropy, purity (mixedness), or several
entanglement witnesses have been widely investigated \cite{Verstraete02,Derkacz,Mazzola10a}.
So far, our knowledge is far from being the simple fact that violating a Bell inequality implies some amount of entanglement.
In particular, for pure states, the presence of entanglement guarantees violation of a Bell inequality
(Gisin's Theorem) \cite{Gisin91}. However, for mixed stares the situation becomes more complicated,
and a great deal of endeavor has been devoted to this subject \cite{Werner89,Munro01,Ghosh01,Miranowicz04}.

On the other hand, quantum discord, proposed as a measure of quantumness of correlations by Ollivier and Zurek \cite{Ollivier01},
and independently by Henderson and Vedral \cite{Henderson01}, has also received much attention from both theoretical \cite{Discord1} and
experimental \cite{Discord2} aspects. Quantum discord is aiming at capturing the nonclassical part of correlations, which
includes entanglement, playing a crucial role in some quantum information processing, especially in the absence
of entanglement \cite{Discord3}. Nevertheless, the analytical expression for quantum discord is only available for
Bell-diagonal states \cite{Luo08} and a certain class of X-structured states \cite{X-state}. The most recent
work \cite{Girolami11} shows that a closed expression for quatum discord of arbitrary two-qubit states cannot be
obtained. Dramatically this computational difficulty in turn motivated the proposals of
alterative measures of quantum correlations, one of which is introduced by Dakic et al. using the Hilbert-Schmidt norm,
called "geometric measure of discord" \cite{Dakic10}. Employing this definition, an analytical formula
was obtained for general two-qubit states in contrast to the original version of discord.

As described above, the relations among nonlocality (Bell violation), mixedness, entanglement have been intensively
studied. Intuitively, Bell violations should also show some connection with quantum discord in some form of
mathematical expressions, since both quantities are put forward to characterize some kind of quantum correlations
contained in bipartite systems from different perspectives. However, the related work can be rarely found
in the literature \cite{Batle11a,Batle11b}. In the present work, we are seeking to address the relationship between geometric
quantum discord and nolocality, quantified by the maximal violation of the Bell inequality in its
CHSH form. From the geometric point of view, we have analytically
proved that the violation of the CHSH inequality for given geometric discord is bounded
by the relation $4\sqrt{D_{G}}\leq B\leq2\sqrt{1+2D_{G}}$ for general Bell diagonal states, which is
in agreement with the numerical result presented in Ref. \cite{Batle11a}. To gain further insights, we have explicitly
illustrated the dynamic picture of Bell violation and geometric discord under several types of decoherence
and our results indicate that these two quantities were closely associated with each other.

The remainder of the paper is arranged as follows. In Sec. II, we give a brief review on the
notation and definitions that will be exploited throughout the paper. In Sec. III, we deduce the exact
lower and upper bounds of Bell violation for given geometric discord. In Sec. IV, we study the dynamical behavior
of these two quantifiers under several different kinds of decoherence processes.
Finally, Sec. V is devoted to the discussion and conclusion.

\section{Preliminaries}
As a starting point, we introduce the notation and definitions that will be employed in the remainder of the paper.
Consider the two-qubit system on the Hilbert space $\mathcal{H}_{AB}=\mathbb{C}^2\otimes\mathbb{C}^2$, the Bell
operator corresponding to Bell-CHSH inequality can be formulated in the following form
\begin{eqnarray}
\mathcal{B}_{CHSH}=\bm{a}\cdot\bm{\sigma}\otimes(\bm{b}+\bm{b'})\cdot\bm{\sigma}
+\bm{a'}\cdot\bm{\sigma}\otimes(\bm{b}-\bm{b'})\cdot\bm{\sigma},
\end{eqnarray}
where $\bm{a}$, $\bm{a'}$, $\bm{b}$, $\bm{b'}$ are the unit vectors in $\mathbb{R}^3$, and $\bm{\sigma}=(\sigma_1,\sigma_2,\sigma_3)$
with $\sigma_1$, $\sigma_2$, $\sigma_3$ being the Pauli matrices. One can write an arbitrary two-qubit state in the Bloch decomposition
\begin{eqnarray}
\label{two-qubit}
\rho=\frac{1}{4}(I\otimes I+\bm{x}\cdot\bm{\sigma}\otimes I+I\otimes\bm{y}\cdot\bm{\sigma}
+\sum_{i,j=1}^{3}t_{ij}\sigma_i\otimes\sigma_j),
\end{eqnarray}
where $\bm{x}$, $\bm{y}$ are vectors in $\mathbb{R}^3$, and $t_{ij}=Tr(\rho\sigma_i\otimes\sigma_j)$ are the real components
of the correlation matrix $T$. Then the well-known CHSH inequality is expressed as
\begin{eqnarray}
\label{CHSH}
B=|\langle\mathcal{B}_{CHSH}\rangle_\rho|=|Tr(\rho\mathcal{B}_{CHSH})|\leq2.
\end{eqnarray}
In Ref. \cite{Horodecki95} the Horodecki family presented the necessary and sufficient condition for violating the CHSH inequality
by an arbitrary two-qubit state.

{\it Lemma 1: (Horodecki's Theorem \cite{Horodecki95}) For any density matrix (\ref{two-qubit}), the maximal violation of the CHSH inequality $\max Tr(\rho\mathcal{B}_{CHSH})$ is given by $2\sqrt{m(\rho)}$, and the inequality (\ref{CHSH}) is violated by some choice of
$\bm{a}$, $\bm{a'}$, $\bm{b}$, $\bm{b'}$ if and only if $m(\rho)>1$, where $m(\rho)=\max_{i<j}(u_i+u_j)$ and $u_i$, $i=1,2,3$
are the eigenvalues of $U=T^{T}T$ ($T^{T}$ denotes the transposition of $T$).}

Meanwhile, geometric measure of quantum discord is defined as \cite{Dakic10}
\begin{equation}
\mathcal{D}_{G}(\rho):=\min_{\chi\in\Omega}\|\rho-\chi\|^2,
\end{equation}
where $\Omega$ denotes the set of zero-discord states and $\|\rho-\chi\|^2=Tr(\rho-\chi)^2$ is the square of
Hilbert-Schmidt norm of Hermitian operators. In the two-qubit case, the geometric measure of quantum discord of Eq. (\ref{two-qubit})
can be evaluated as \cite{Dakic10}
\begin{equation}
\mathcal{D}_{G}(\rho)=\frac{1}{4}(\|\vec{x}\|^2+\|T\|-k_{max}).
\end{equation}
where $k_{max}$ is the largest eigenvalue of matrix $K=\vec{x}\vec{x}^T+TT^T$.

\section{Bell violation versus geometric measure of quantum discord}
In this section, we restrict our attention to a specific class of states, Bell diagonal states, not only
because of analytical simplicity but also their important role in entanglement concentration \cite{Bennett96},
local filtering operations \cite{Verstraete02}, and security proof of DIQKD \cite{Pironio09} et al. In Bloch representation,
Bell diagonal states can be expressed as
\begin{align}
\label{Bell}
\rho_{Bell}=&\frac{1}{4}(I\otimes I+\sum_{i=1}^{3}c_{i}\sigma_i\otimes\sigma_i)\\
=&\lambda_{\Phi}^{+}|\Phi^{+}\rangle\langle\Phi^{+}|+\lambda_{\Psi}^{+}|\Psi^{+}\rangle\langle\Psi^{+}|\nonumber\\
&+\lambda_{\Phi}^{-}|\Phi^{-}\rangle\langle\Phi^{-}|+\lambda_{\Psi}^{-}|\Psi^{-}\rangle\langle\Psi^{-}|
\end{align}
where $|\Psi^{\pm}\rangle=\frac{1}{\sqrt{2}}(|01\rangle\pm|10\rangle)$, $|\Phi^{\pm}\rangle=\frac{1}{\sqrt{2}}(|00\rangle\pm|11\rangle)$
are the four Bell states, and $c_i$ are real constants fulfilling certain constraints such that $\rho_{Bell}$ is positive
semi-definite, that is
\begin{subequations}
\label{positivity}
\begin{align}
& 0 \leq \lambda_{\Psi}^{-}=\frac{1}{4} (1 - c_{1} - c_{2} - c_{3}) \leq 1, \label{positivity1}\\
& 0 \leq \lambda_{\Phi}^{-}=\frac{1}{4} (1 - c_{1} + c_{2} + c_{3}) \leq 1, \label{positivity2}\\
& 0 \leq \lambda_{\Phi}^{+}=\frac{1}{4} (1 + c_{1} - c_{2} + c_{3}) \leq 1, \label{positivity3}\\
& 0 \leq \lambda_{\Psi}^{+}=\frac{1}{4} (1 + c_{1} + c_{2} - c_{3}) \leq 1, \label{positivity4}
\end{align}
\end{subequations}
Utilizing these conditions, we present our Lemma 2 as a geometric representation of Bell diagonal states.

{\it Lemma 2: (Geometrical representation) If a Bell diagonal state is physical, then the following
restrictions must be satisfied: $|c_{i}|\leq1$ for $i=1,2,3$ and $c_{i}^2-c_{j}^2-c_{k}^2\geq-1$ with $i\neq j\neq k$.}

Proof: Form inequalities $(\ref{positivity1})+(\ref{positivity2})$ and $(\ref{positivity3})+(\ref{positivity4})$,
we can easily find
\begin{eqnarray}
-3\leq c_1\leq1,\quad -1\leq c_1\leq3 \quad \Rightarrow |c_{1}|\leq1,
\end{eqnarray}
Besides, from $(\ref{positivity1}) \times (\ref{positivity4})$ and $(\ref{positivity2}) \times (\ref{positivity3})$,
we have
\begin{subequations}
\label{constraint}
\begin{align}
& (1-c_{3})^2 - (c_{1}+c_{2})^2 \geq0, \label{constraint1}\\
& (1+c_{3})^2 - (c_{1}-c_{2})^2 \geq0, \label{constraint2}
\end{align}
\end{subequations}
Consequently from $(\ref{constraint1})+(\ref{constraint2})$, we can obtain
\begin{eqnarray}
c_{3}^2-c_{1}^2-c_{2}^2\geq-1.
\end{eqnarray}
Similarly, other constraints can be verified: $|c_{2}|\leq1$, $|c_{3}|\leq1$ and $c_{1}^2-c_{2}^2-c_{3}^2\geq-1$,
$c_{2}^2-c_{1}^2-c_{3}^2\geq-1$ due to symmetry property of the 3-tuple $(c_1,c_2,c_3)$.

In fact, in Ref \cite{Horodecki96} the Horodecki family already advanced the geometrical representation of Bell diagonal
states, which depicts a tetrahedron with vertices $(1,1,-1)$, $(-1,-1,-1)$,  $(1,-1,1)$ and  $(-1,1,1)$.
Recently, this geometric interpretation has been expanded in some other context \cite{Spengler11}. Before presenting
our main result, we first consider the relationship between $B$ and $D_{G}$ for pure states since they
possess the maximum nonlocality in the sense that "entanglement" just indicates "Bell violations"
in sharp contrast to mixed states \cite {Gisin91}. Up to local unitary equivalence, any pure two-qubit state
can be written as
\begin{eqnarray}
|\theta\rangle=\cos\theta \ |00\rangle+\sin\theta \ |11\rangle,
\end{eqnarray}
To calculate Bell violation and geometric discord of pure states, it is convenient to decompose
$\rho=|\theta\rangle\langle\theta|$ in Bloch form
\begin{align}
\rho=&\frac{1}{4}(I+\cos2\theta \ \sigma_z\otimes I+\cos2\theta \ I\otimes \sigma_z\nonumber\\
&+\sum_{i=1}^3c_i\sigma_i\otimes\sigma_i)
\end{align}
with the correlation components
\begin{eqnarray}
c_1=\sin2\theta, \ c_2=-\sin2\theta, \ c_3=1.
\end{eqnarray}
It is straightforward to obtain
\begin{align}
B=&2\sqrt{1+\sin^{2}2\theta}, \\
D_{G}=&\frac{1}{4}(c_1^2+c_2^2+c_3^2+\cos^22\theta\nonumber\\
&-\max\{c_1^2,c_2^2,c_3^2+\cos^22\theta\})\nonumber\\
=&\frac{1}{2}\sin^{2}2\theta,
\end{align}
where $c=\max(c_1,c_2,c_3)$. Obviously, we arrive at the relation
\begin{eqnarray}
B=2\sqrt{1+2D_{G}}.
\end{eqnarray}
As described above, it is reasonable to conjecture that this relationship holds as upper bound for all Bell diagonal states.
Now we can formulate the main theorem of this paper.

\begin{figure}[htbp]
\begin{center}
\resizebox{0.4\textwidth}{!}{%
\includegraphics{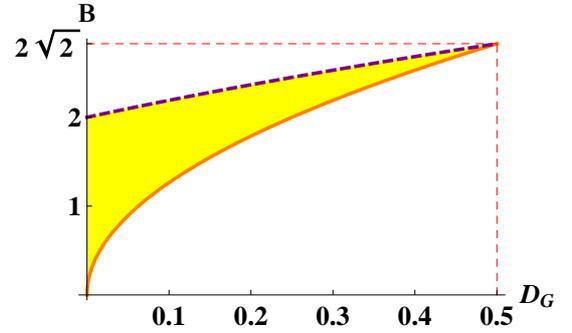}
}
\end{center}
\caption{(Color online) The region of possible maximal Bell violation versus geometric discord for Bell diagonal states.
The orange solid curve represents Werner states while purple dashed curve stands for rank 1 or 2 Bell diagonal states.
}\label{bounds}
\end{figure}

{\it Theorem: For Bell diagonal states, the violation of the CHSH inequality for given geometric discord is bounded
by the relation: $4\sqrt{D_{G}}\leq B\leq2\sqrt{1+2D_{G}}$(see Fig. \ref{bounds}).}

Proof: 1)The lower bound. Using Lemma 1, for Bell diagonal states (\ref{Bell}), CHSH violation is simply given by
\begin{gather}
B=2\sqrt{m(\rho)},\nonumber\\
\ m(\rho)=\max\{c_1^2+c_2^2,c_2^2+c_3^2,c_3^2+c_1^2\},
\end{gather}
On the other hand, geometric measure of discord can be evaluated as \cite{Dakic10}
\begin{align}
D_{G}&=\frac{1}{4}(c_1^2+c_2^2+c_3^2-c^2)\nonumber\\
&=\frac{1}{4}\min\{c_1^2+c_2^2,c_2^2+c_3^2,c_3^2+c_1^2\},
\end{align}
It is easy to see that $B\geq4\sqrt{D_{G}}$.

2)The upper bound. Without loss of any generality, one can assume $c_1^2\geq c_2^2\geq c_3^2$,
and then check the quantity
\begin{align}
\label{difference}
\Delta &=1+2D_{G}-\frac{1}{4}B^2\nonumber\\
&=1+\frac{1}{2}(c_2^2+c_3^2)-(c_1^2+c_2^2)\nonumber\\
&=1+\frac{1}{2}(c_3^3-2c_1^2-c_2^2)\nonumber\\
&\geq1+\frac{1}{2}(-1-1) \qquad (using \ Lemma \ 2)\nonumber\\
&=0.
\end{align}
Note that in the proof we have employed the constraints $-c_1^2\geq-1$ and $c_{3}^2-c_{1}^2-c_{2}^2\geq-1$.
Inequality (\ref{difference}) is just equivalent to $B\leq2\sqrt{1+2D_{G}}$, which is just the upper bound as we
expect.

Furthermore, we notice that the lower bound is satisfied by Werner states \cite{Werner89}
\begin{eqnarray}
\label{Werner}
\rho_{W}=c|\Psi^{-}\rangle\langle\Psi^{-}|+(1-c)\frac{I}{4},\quad c\in[0,1]
\end{eqnarray}
Meanwhile, the upper bound is attained for rank 1 or 2 Bell diagonal states ($c_1=\pm1$ and $c_2=\mp c_3$)
\begin{eqnarray}
\label{upper}
\rho=\frac{1+c_3}{2}|\Phi^{\pm}\rangle\langle\Phi^{\pm}|+\frac{1-c_3}{2}|\Psi^{\pm}\rangle\langle\Psi^{\pm}|,
\end{eqnarray}
It is interesting to note that these states (\ref{upper}) are just states which play an essential part in sudden transition
between classical and quantum decoherence \cite{Mazzola10b}. In Ref. \cite{Verstraete02}, Verstraete and Wolf presented the maximal violation and minimal violation of the CHSH inequality for given concurrence $C$: $\max(2,2\sqrt{2}C)\leq B\leq2\sqrt{1+C^2}$, and
specially for Bell diagonal states, the region of possible violations becomes
\begin{eqnarray}
\frac{2\sqrt{2}(2C+1)}{3}\leq B\leq2\sqrt{1+C^2}.
\end{eqnarray}
where the lower and upper bounds are sharp for Werner states and rank 2 Bell diagonal states respectively,
which is remarkably consistent with our results. Note that for Werner states (\ref{Werner}),
the concurrence $C(\rho)=\max\{0,(3c-1)/2\}$, and correspondingly geometric discord
$D_{G}=(c^2+c^2+c^2-c^2)/4=c^2/2$, such that we have the relation $4\sqrt{D_{G}}=2\sqrt{2}(2C(\rho)+1)/3$ (if $c\geq1/3$).

\section{Dynamical behavior}
In this section, we investigate the connection between Bell violation and geometric measure of discord in dynamical
context. Quantum discord, as a measure of correlations of bipartite systems, will inevitably decohere due to the
interaction between systems and environment. Both theory and experiment have been put forward to examine the dynamics
of quantum discord under Markovian \cite{Maziero09,Werlang09} and non-Markovian \cite{Fanchini09,Wang10} environments.
It is remarkable that quantum discord always shows a certain degree of robustness against decoherence
where in same cases entanglement may display the phenomenon of "sudden death".
Besides, the behaviors of geometric version of quantum discord under various decoherence have also
been studied in the literature \cite{Lu10,Xu10}. Recently, a number of papers have been devoted to the analysis of effects
of environmental decoherence on the Bell-CHSH inequality violation for qubit systems as well \cite{Bell-decoherence}. However,
these two quantities have rarely been considered under the same dynamic conditions. In this work we focus on
the comparison of their performance under several typical decoherence processes so as to verify whether they
are related to each other in a particular way.
\subsection{Markovian noise}
\begin{figure}[htbp]
\begin{center}
\subfigure[]{
\label{a}
\includegraphics[width=0.2\textwidth ]{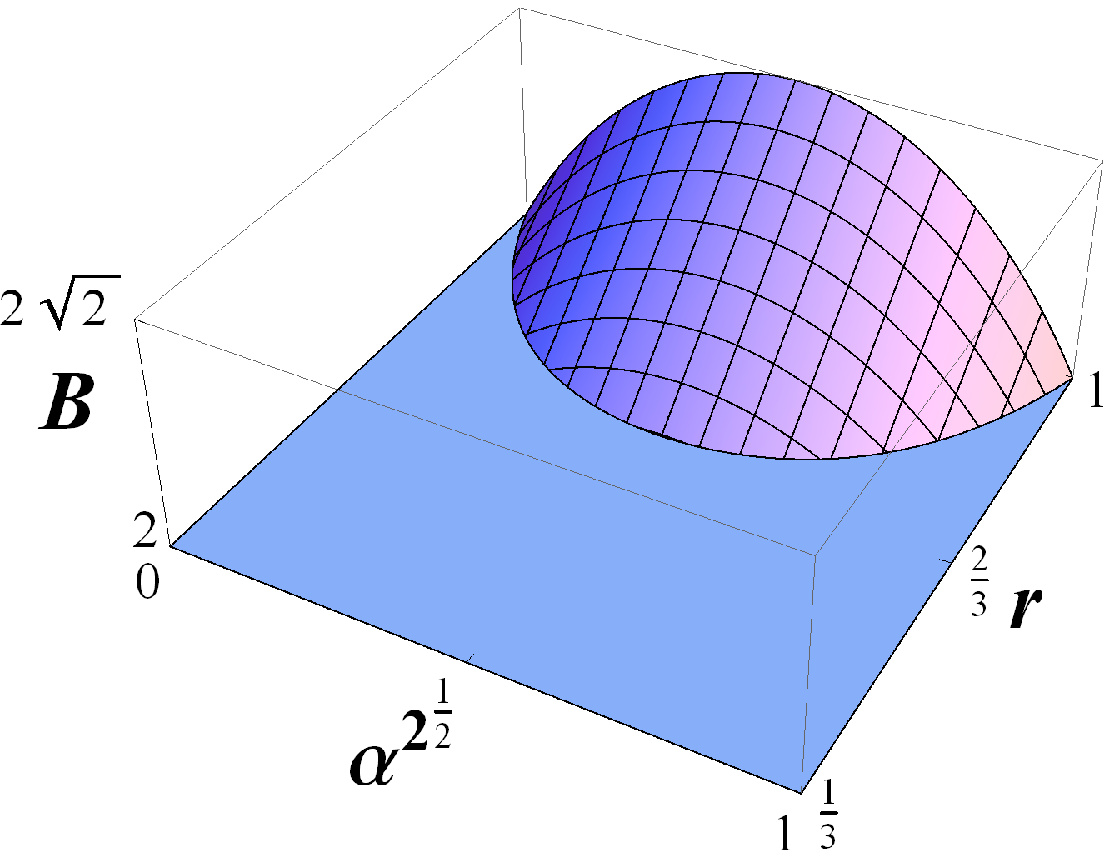}}
\subfigure[]{
\label{b}
\includegraphics[width=0.2\textwidth ]{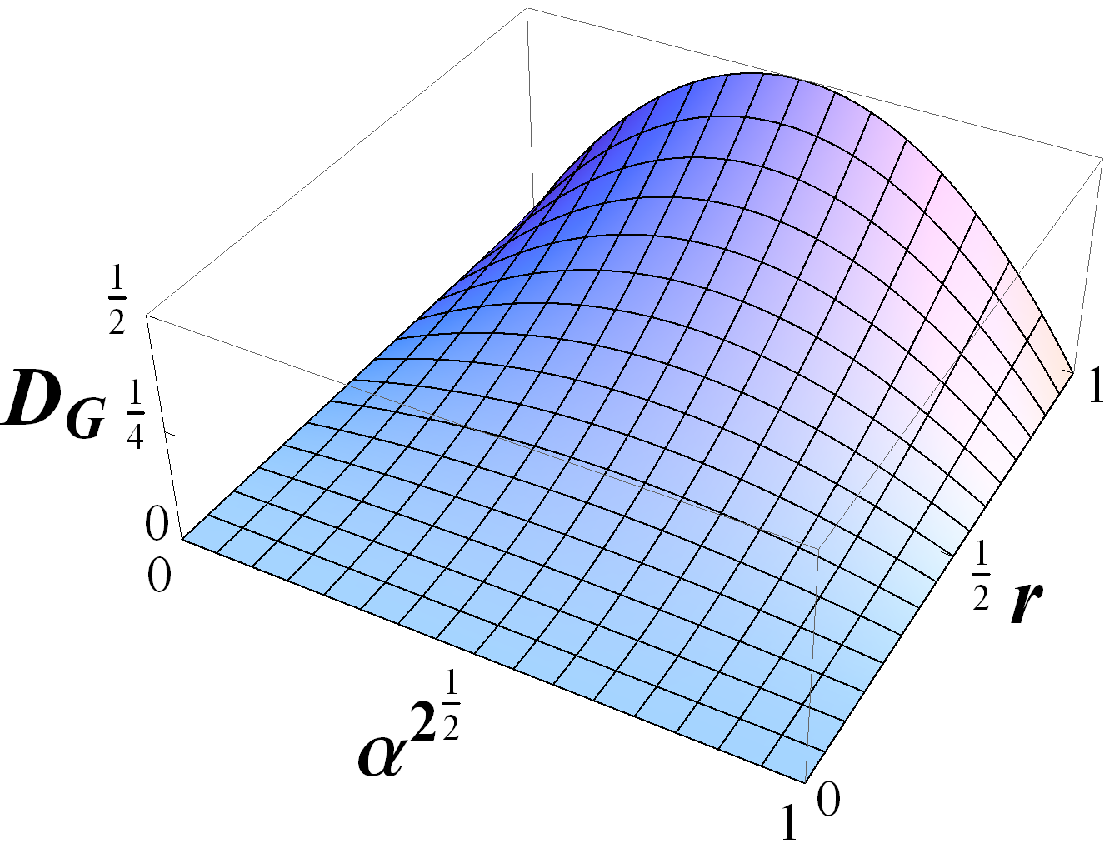}}
\end{center}
\caption{(Color online) Bell violation $B$ and geometric discord $D_G$ for initial EWL states
as a function of parameters $\alpha^2$ and $r$.
}\label{initial}
\end{figure}

We first take the phase damping (or equivalently phase flip) channel for example. Following the Kraus operator approach,
the evolved state of a two-qubit system $\rho_{AB}$ under local environments can be modeled in the Kraus representation \cite{Kraus}
\begin{equation}
\label{Kraus}
\varepsilon(\rho)=\sum_{i,j}K_{i,j}\rho(0)K_{i,j}^{\dag},
\end{equation}
where $K_{i,j}=K^A_i\otimes K^B_j$ are Kraus operators, satisfying $K_{i,j}^{\dag}K_{i,j}=I$ if the quantum operation
is trace-preserving, and the operators $K_{i(j)}$ denote the single-qubit decoherence effects. For phase damping channel,
the Kraus operators are given by
\begin{align}
\label{phase-damping}
K_0 = \left(\begin{array}{cc}
1 & 0 \\
0 & \sqrt{1-p}
\end{array}\right), \,
K_1 = \left(\begin{array}{cc}
0 & 0 \\
0 & \sqrt{p}
\end{array}\right)
\end{align}
where the parametrized time p is responsible for a wide range of physical phenomena \cite{Nielsen}. Here, we restrict
our consideration to extended Werner-like (EWL) two-qubit initial states
\begin{align}
\label{EWL}
& \rho^\Phi=r|\Phi\rangle\langle\Phi|+\frac{1-r}{4}I_4,\nonumber\\
& \rho^\Psi=r|\Psi\rangle\langle\Psi|+\frac{1-r}{4}I_4,
\end{align}
where $r$ quantifies the purity of EWL states, and when $r=1$ EWL states reduce to Bell-like pure states
\begin{align}
|\Phi\rangle=\alpha|01\rangle+\beta|10\rangle,\
|\Psi\rangle=\alpha|00\rangle+\beta|11\rangle,
\end{align}
with real parameters $\alpha$ and $\beta$ defining the degree of entanglement of pure part of EWL states
and $\alpha^2+\beta^2=1$. Employing concurrence  \cite{Wootters} as the measure of entanglement, the initial entanglement
for both EWL states of Eq. (\ref{EWL}) is the same, given by
\begin{align}
C^\Phi=C^\Psi=2\max\{0,(\alpha\beta+1/4)r-1/4\},
\end{align}
Therefore it is easy to find that the initial EWL states are entangled only if $r>(1+4\alpha\beta)^{-1}\geq 1/3$.
We plot the initial Bell violation and geometric discord in Fig. \ref{initial}.
\begin{figure}[htbp]
\begin{center}
\subfigure[]{
\label{a}
\includegraphics[width=0.2\textwidth ]{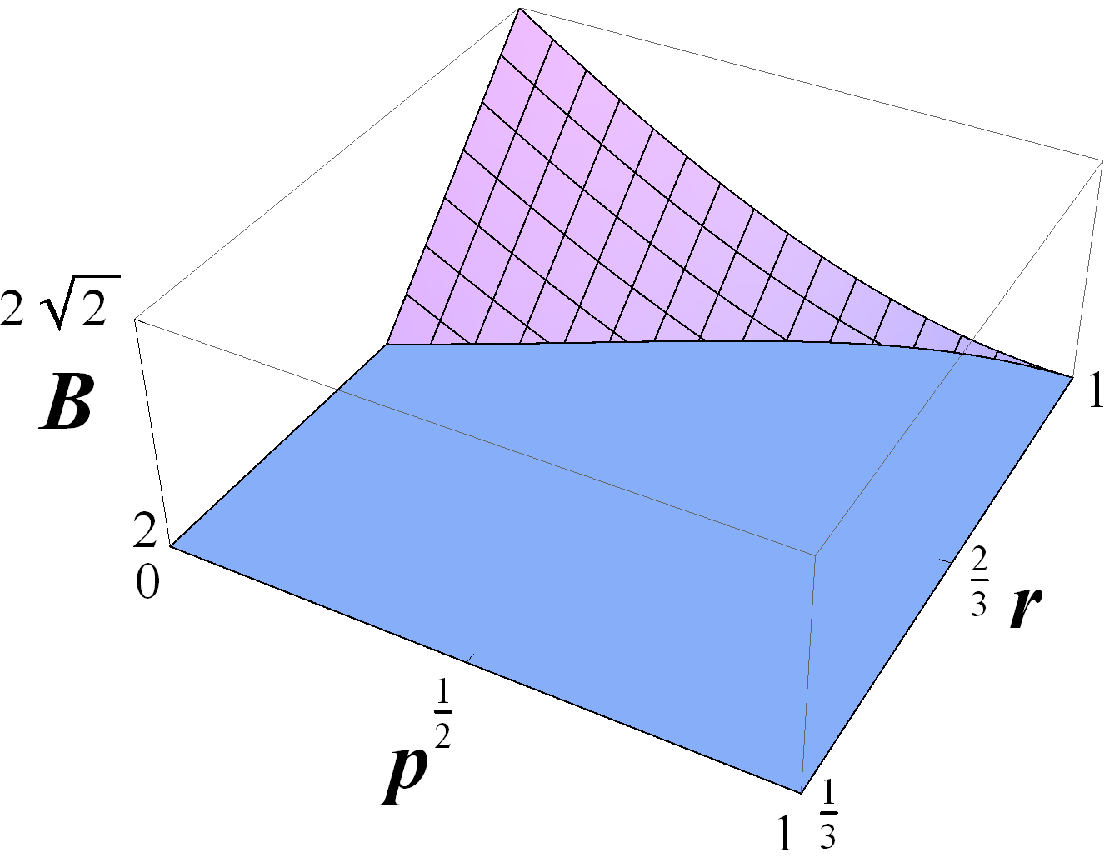}}
\subfigure[]{
\label{b}
\includegraphics[width=0.2\textwidth ]{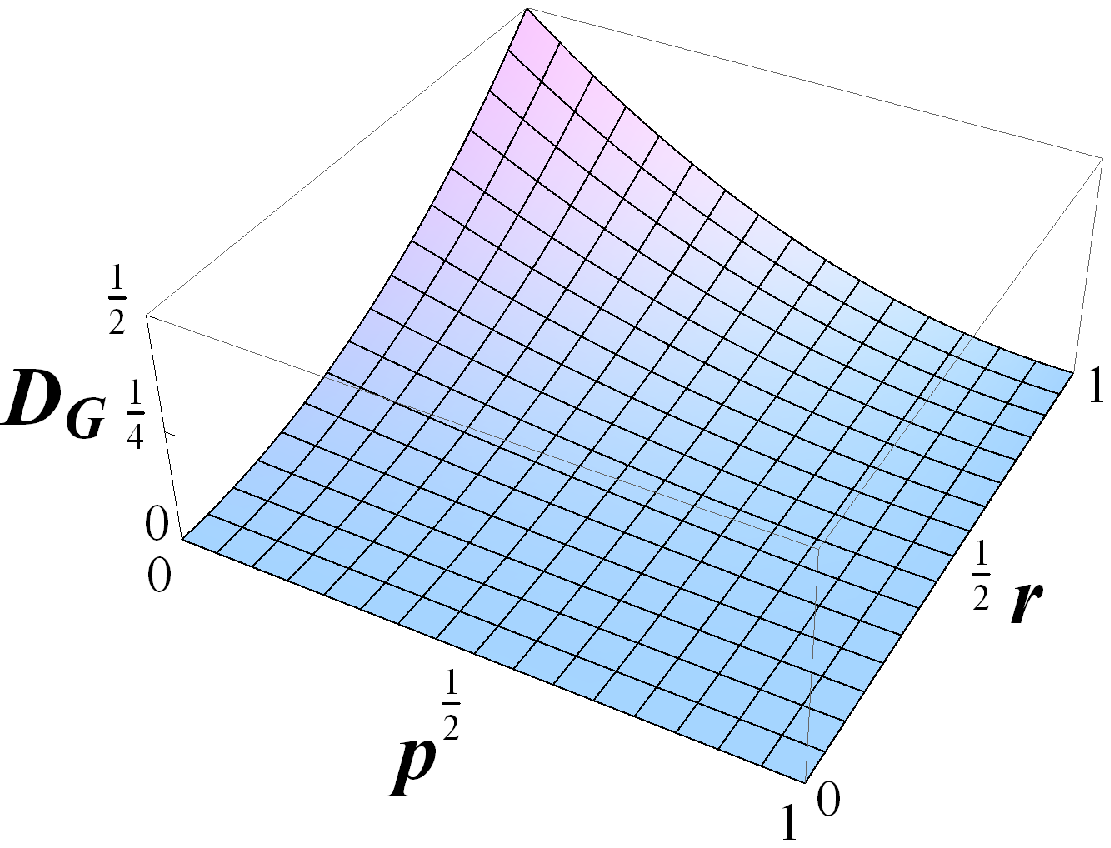}}\\
\subfigure[]{
\label{c}
\includegraphics[width=0.2\textwidth ]{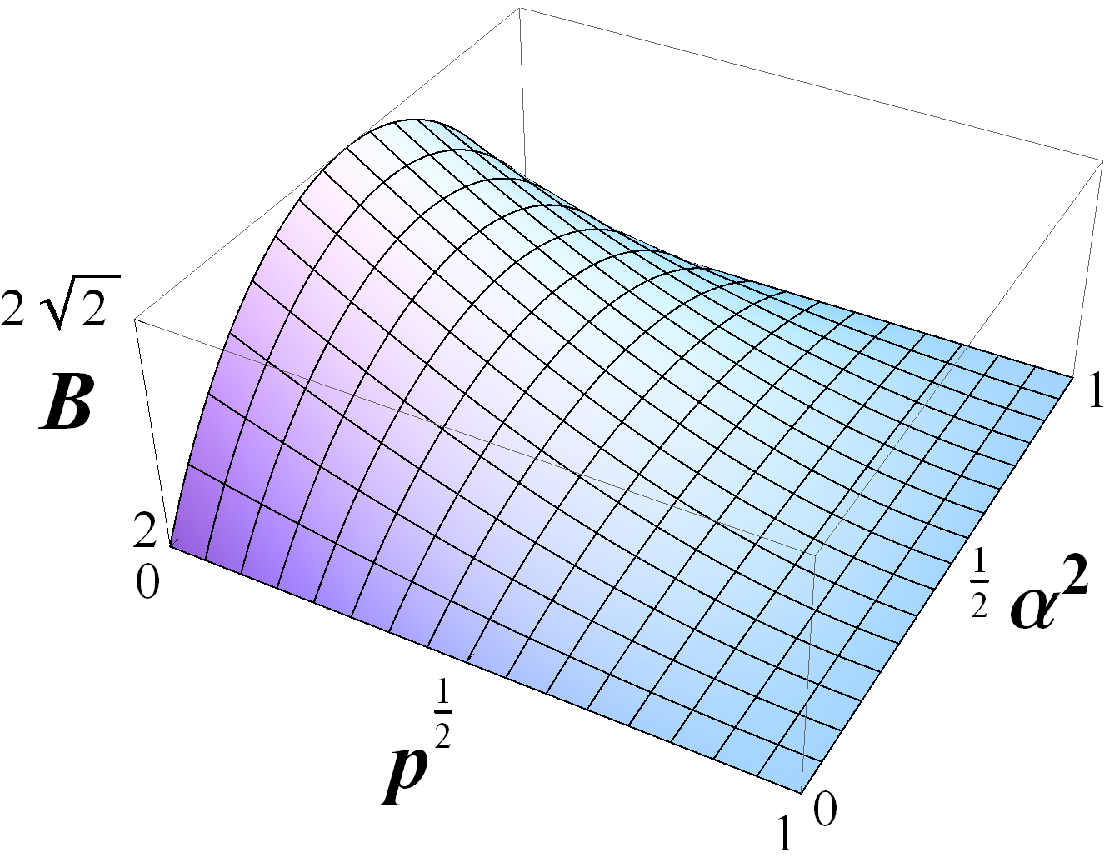}}
\subfigure[]{
\label{d}
\includegraphics[width=0.2\textwidth ]{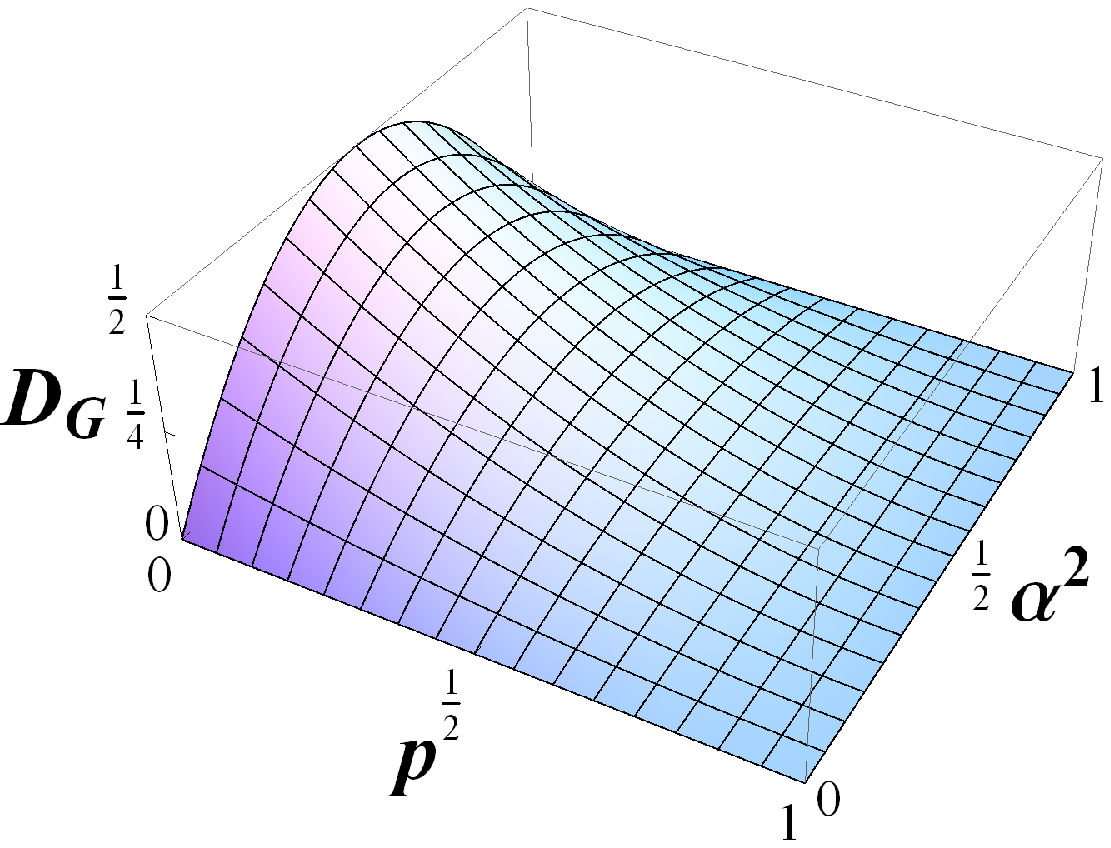}}
\end{center}
\caption{(Color online) Bell violation $B$ and geometric discord $D_G$ for decohered EWL states
as a function of various parameters : (a)(b) $\alpha^2=\beta^2=1/2$ (Werner-like states); (c)(d) $r=1$
(Bell-like pure states).
}\label{PD}
\end{figure}

In Appendix (\ref{Markovian}), we obtain the explicit expressions of $B$ and $D_G$ for EWL states
under phase damping channel. In Fig. \ref{PD}, we show the effect of phase damping on EWL states
for different parameters. For Werner-like states ($\alpha^2=\beta^2=1/2$), the "violation
sudden death" \cite{Bellomo08,Jaeger08} clearly arises in contrast to the well-known phenomenon "entanglement sudden death"
which makes its appearance in a wide range of physical situations \cite{Yu}. On the contrary, geometric measure
of discord shows its robustness against decoherence behaving similarly to the original version of discord \cite{Werlang09}.
For Bell-like states ($r=1$), we note that the dynamics of $B$ and $D_G$ are almost the same, and the maintenance
of Bell nonlocality ($B>2$) is more sensitive to the initial purity $r$ than $\alpha^2$ and symmetric with respect
to $\alpha^2=1/2$. Moreover, it is remarkable to see that the relation $B=2\sqrt{r^2+2D_G}$ always holds for both initial
and decohered EWL states (see appendix for more details).
\subsection{non-Markovian noise}
In this subsection, we analyze the non-Markovian dynamics of Bell violation and geometric discord
concerning two-qubit system, each locally and independently interacting with a zero-temperature resevoir. The
exact solution of single-qubit dynamics can be found in Ref. \cite{Bellomo07}, where the physical process is in fact
modeled as the amplitude decay channel
\begin{align}
\rho(t) = \left(\begin{array}{cc}
\rho_{11}(0)P_t & \rho_{10}(0)\sqrt{P_t} \\
\rho_{01}(0)\sqrt{P_t} & \rho_{00}(0)+\rho_{11}(0)(1-P_t)
\end{array}\right),
\end{align}
and the function $P_t$ takes the form
\begin{align}
P_t=e^{-\lambda t}\left[\cos(\frac{dt}{2})+\frac{\lambda}{d}\sin(\frac{dt}{2})\right]^2,
\end{align}
where $d=\sqrt{2\Gamma\lambda-\lambda^2}$, and the parameter $\lambda$ denotes the spectral width of the
coupling, $\Gamma$ represents the decay rate of the excited state in the Markovian limit of a flat spectrum.
\begin{figure}[htbp]
\begin{center}
\subfigure[]{
\label{a}
\includegraphics[width=0.2\textwidth ]{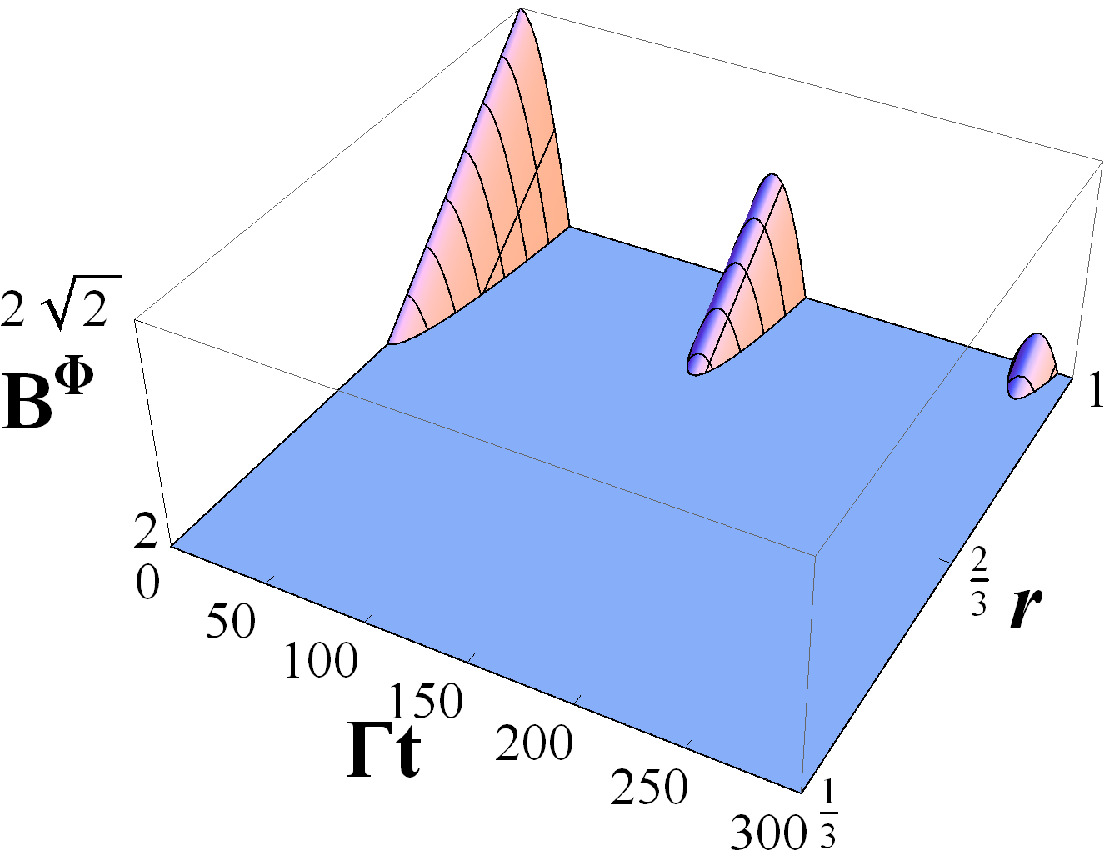}}
\subfigure[]{
\label{b}
\includegraphics[width=0.2\textwidth ]{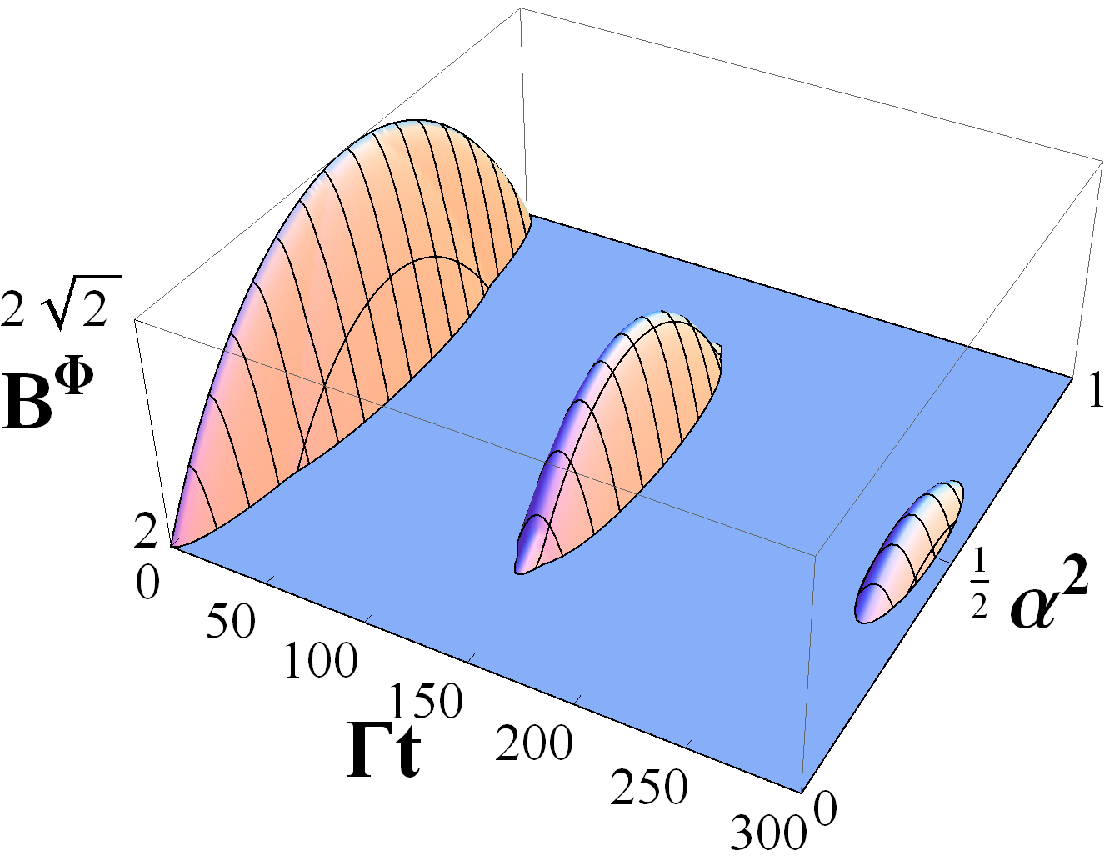}}\\
\subfigure[]{
\label{c}
\includegraphics[width=0.2\textwidth ]{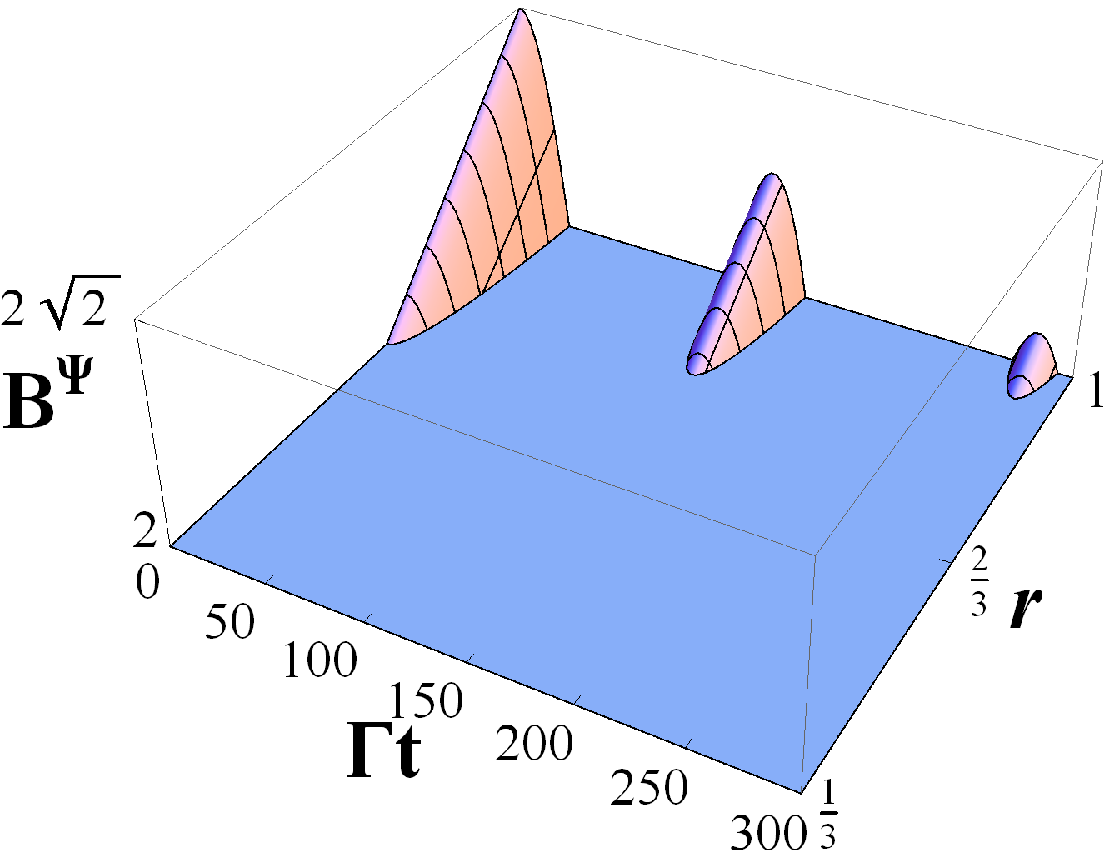}}
\subfigure[]{
\label{d}
\includegraphics[width=0.2\textwidth ]{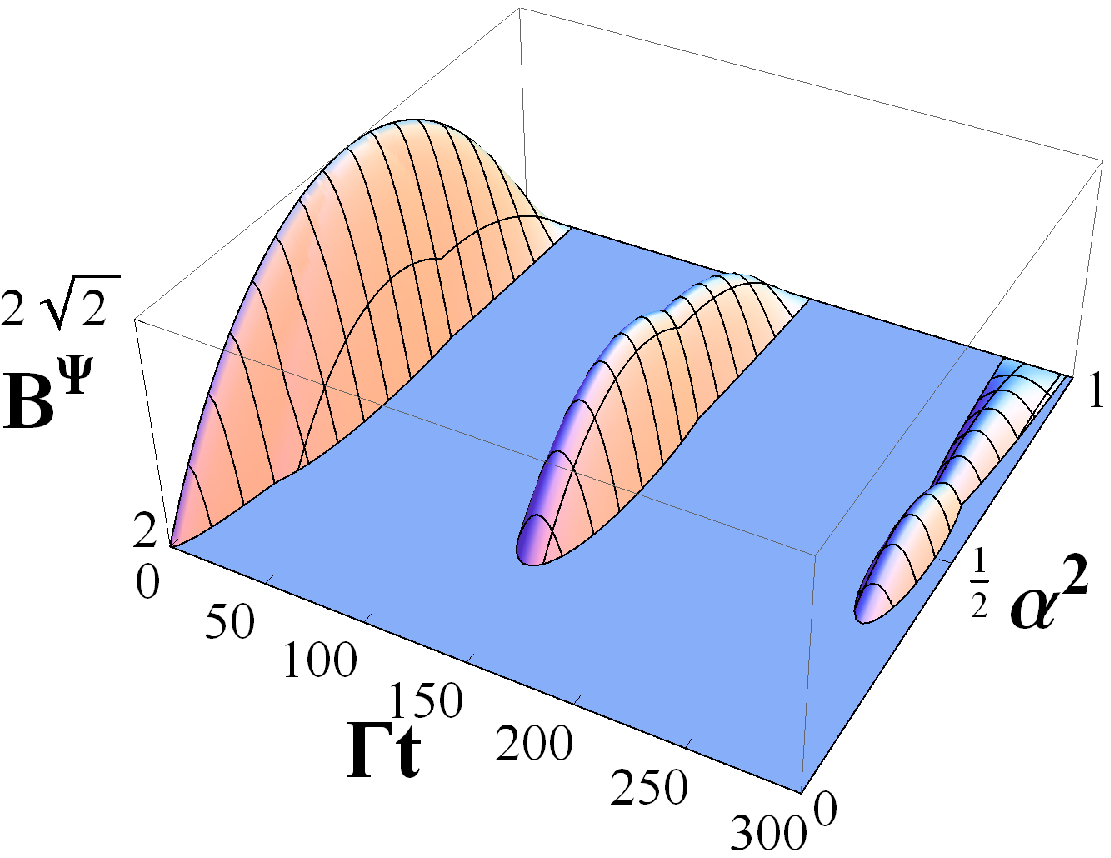}}
\end{center}
\caption{(Color online) The dynamics of Bell violation $B^\Phi$ ((a)(b)) and $B^\Phi$ ((c)(d))
in non-Markovian case as a function of the dimensionless quantity $\Gamma t$, $r$, or $\alpha^2$ ($\lambda/\Gamma=10^{-3}$):
(a)(c) $\alpha^2=\beta^2=1/2$ (Werner-like states); (b)(d) $r=1$ (Bell-like pure states).
}\label{AD}
\end{figure}

Exploiting the calculations in Appendix (\ref{non-Markovian}), it is convenient to implement
any dynamical simulation with respect to parameters $\Gamma t$, $r$, or $\alpha^2$. For instance,
we can easily retrieve and extend the results in Ref. \cite{Bellomo08}.  From Fig. \ref{AD}, we see that
"violation sudden death" and "violation sudden birth" or so called "Bell islands" \cite{Bellomo08} occurs
for both initial Werner-like states and Bell-like states. For Werner-like states, the dynamics of
$\rho^\Phi$ and $\rho^\Psi$ is just the same; however, for Bell-like states, the dynamics of
$\rho^\Phi$ is symmetric with respect to $\alpha^2=1/2$ while $\rho^\Psi$ is not. Note that the
sudden death and revival of Bell violation is closely related to the memory effect of non-Markovian
environment. If the non-Markovian effect is not sufficiently strong, for example, $\lambda/\Gamma$
is larger than $10^{-2}$, thus the revival can never appears.
\begin{figure}[htbp]
\begin{center}
\subfigure[]{
\label{a}
\includegraphics[width=0.2\textwidth ]{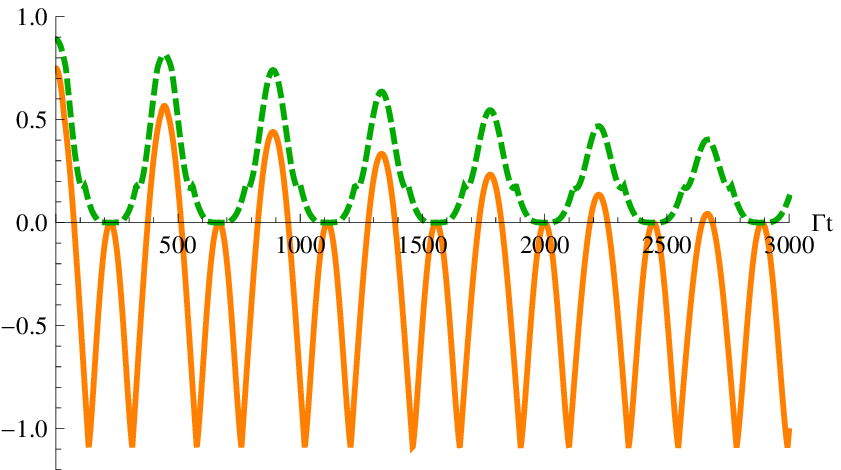}}
\subfigure[]{
\label{b}
\includegraphics[width=0.2\textwidth ]{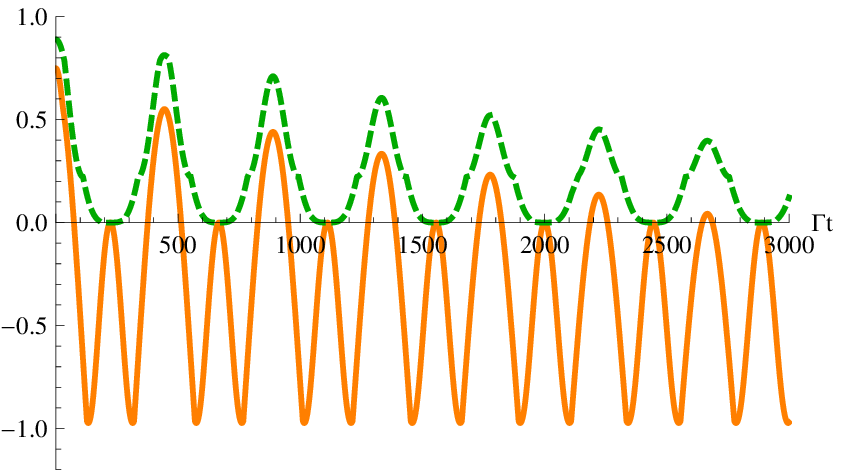}}\\
\subfigure[]{
\label{c}
\includegraphics[width=0.2\textwidth ]{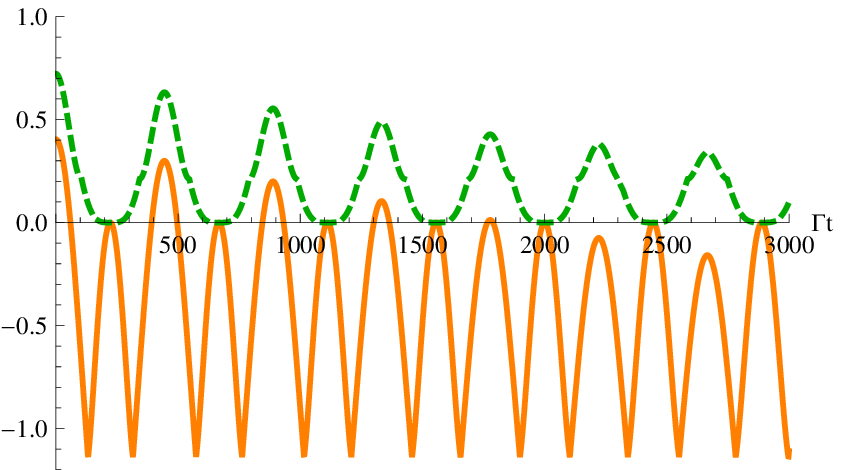}}
\subfigure[]{
\label{d}
\includegraphics[width=0.2\textwidth ]{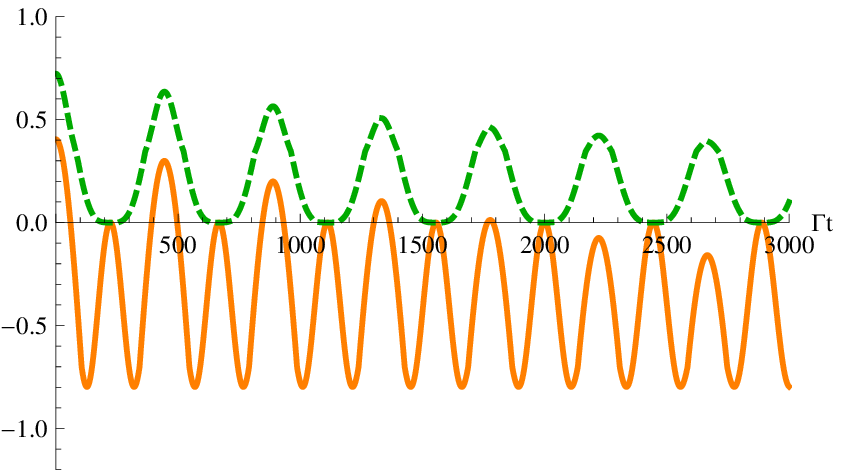}}
\end{center}
\caption{(Color online) Bell violation $B-2$ (orange solid line) and normalized geometric discord $2D_G$ (green dashed line)
in non-Markovian case ($\lambda/\Gamma=10^{-4}$) as a function of the dimensionless quantity $\Gamma t$ for $\rho^\Phi$ ((a)(c))
and $\rho^\Psi$ ((b)(d)): (a)(b) $\alpha^2=1/3$, $r=1$ ; (c)(d) $\alpha^2=1/2$, $r=0.85$.
}\label{comparison}
\end{figure}

Futhermore, for comparison, we plot the time evolution of the Bell violation $B-2$ and normalized geometric
discord $2D_G$ (the maximum value of $D_G$ is $1/2$ for two-qubit states) for the same fixed value of parameters
in Fig. \ref{comparison}. From Ref. \cite{Bellomo07}, we know that the function $P_t$ only has discrete zeros at
$t_n=2[n\pi-\arctan(d/\lambda)]/d$ where $n$ is integer. The plot clearly shows the quantity $B-2$ and $2D_G$
act similarly along with the function $P_t$: simultaneously and periodically reach zero according to the zero points
of $P_t$ and achieve extreme values at the maximum points of $P_t$, no matter which states we consider, $\rho^\Phi$ or $\rho^\Psi$
(one can easily obtain the exact expressions of $B-2$ and $2D_G$ from Appendix (\ref{non-Markovian}) so as to justify
all the details). These figures show that times of revival of Bell violation sensibly depend on the purity parameter $r$,
and we see that times of revival reduce if $r$ becomes smaller. However, geometric discord only vanishes at discrete points
and still remains robust against non-Markovian noise like the situation in the Markovian case.
\subsection{non-back-action environment}
As has been mentioned above, the occurrence of Bell violation revivals originates from the memory
effect of the non-Markovian quantum reservoirs, or more precisely, due to the back-action of qubit
system on the corresponding environment. This back-action mechanism leads to establishment of quantum correlations
between qubits and environments or between the environments themselves. Here, we investigate the case where a pair
of independent qubits each locally subject to the so called "system-unaffected environment" (SUE) \cite{Franco10}, which means
the back-action of system does not exist, in order to see whether the revivals of Bell violation or geometric discord
would happen in this model.

Consider a initial two-qubit system $\rho(0)$ each locally coupled to random external fields, which can be
characterized as a bistochastic quantum channel \cite{Zyczkowski01}. The dynamical map $\Lambda$ can be always written as \cite{Alicki}
\begin{align}
\Lambda[\rho(0)]=\sum_{i=1}^{N}p_iA_i\rho(0)A_i^\dag,
\end{align}
where $p_i$ is a probability measure with $\Sigma_{i=1}^{N}p_i=1$, $p_i\geq0$ and $A_i$, $i=1,2,...,N$ are unitary operators.
The environment depicted in Ref. \cite{Franco10} can be imagined as follows: two classical external fields have the same frequency and amplitude but each pass through a random dephaser which switches between the values $0$ and $\pi$ with the same probability,
and then locally act on one of the qubit pair. For this case the single-qubit dynamical map can be written as $\Lambda_S[\rho_S(0)]=\frac{1}{2}\sum_{i=1}^2U_i^S(t)\rho_S(0)U_i^{S\dag}(t)$, where $U_i^S(t)=e^{-iH_it/\hbar}$ is the time evolution
operator with $H_i=\hbar\omega\sigma_{z}/2+g(\sigma_+e^{-i\phi_i}+\sigma_-e^{i\phi_i})$ and can be represented in matrix form in $\{|1\rangle,|0\rangle\}$ basis
\begin{align}
U_i^S(t) = \left(\begin{array}{cc}
\cos(gt) & -e^{-i\phi_i}\sin(gt) \\
e^{i\phi_i}\sin(gt) & \cos(gt)
\end{array}\right),
\end{align}
Successively, the time behavior of the two-qubit system $\rho_{AB}$ can be given by
\begin{align}
\rho(t)=\Lambda[\rho(0)]=\frac{1}{4}\sum_{i,j=1}^2U_i^A(t)U_j^B(t)\rho(0)U_i^{A\dag}U_j^{B\dag},
\end{align}
A straightforward calculation then shows that the map $\Lambda$ works inside the class of Bell-diagonal states,
and particularly the effect of this map on Bell states have the form
\begin{align}
\Lambda|\beta_\pm\rangle\langle \beta_\pm|=[1-f(t)]|\beta_\pm\rangle\langle \beta_\pm|+f(t)|\beta'_\mp\rangle\langle \beta'_\mp|,
\end{align}
where $\beta,\beta'=\Psi,\Phi$ with $\beta\neq \beta'$ and $|\Psi_\pm\rangle=(|01\rangle\pm|10\rangle)/\sqrt{2}$,
$|\Phi_\pm\rangle=(|00\rangle\pm|11\rangle)/\sqrt{2}$, and $f(t)=\sin^2(2gt)/2$. For a general Bell-diagonal state
$\rho=\sum_{\beta,s}\lambda_\beta^s(0)|\beta_s\rangle\langle \beta_s|$ ($\beta=\Psi,\Phi$; $s=\pm$), the time-dependent
coefficients of $\rho(t)$ can be obtained as
\begin{align}
\label{coefficient}
\lambda_\beta^{\pm}(t)=\lambda_\beta^{\pm}(0)[1-f(t)]+\lambda_{\beta'}^{\mp}(0)f(t),\ (\beta\neq \beta')
\end{align}

\begin{figure}[htbp]
\begin{center}
\includegraphics[width=.40\textwidth]{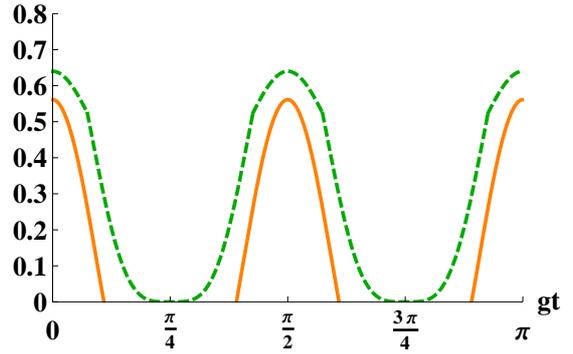} {}
\end{center}
\caption{(Color online) The time behavior of Bell violation $B-2$ (orange solid line) and normalized geometric discord
$2D_G$ (green dashed line) as a function of the dimensionless quantity $gt$ for an initial Bell-diagonal states
$\lambda_1^+(0)=0.9$, $\lambda_1^-(0)=0.1$.
}\label{non-back-action}
\end{figure}
Though it is easy to give the explicit formula for Bell violation or geometric discord based on all the unspecified
parameters $\lambda_\beta^{\pm}(0)$, we now focus on the initial conditions $\lambda_\Psi^+(0)=0.9$, $\lambda_\Psi^-(0)=0.1$,
and $\lambda_\Phi^+(0)=\lambda_\Phi^-(0)=0$. Utilizing Eq. (\ref{coefficient}), we have
\begin{gather}
\lambda_\Psi^+=0.9[1-f(t)],\lambda_\Psi^-=0.1[1-f(t)],\nonumber\\
\lambda_\Phi^+=0.1f(t),\lambda_\Phi^-=0.9f(t),
\end{gather}
Note that for Bell-diagonal states in Bell basis, the maximum possible violation reads as \cite{Verstraete02,Miranowicz04}
\begin{align}
B=2\sqrt{2}\sqrt{(\lambda_1-\lambda_4)^2+(\lambda_2-\lambda_3)^2},
\end{align}
with eigenvalues $\lambda_1\geq\lambda_2\geq\lambda_3\geq\lambda_4$. In our case, $\lambda_\Psi^+$ is largest
and $\lambda_\Phi^+$ is smallest among the eigenvalues, due to $f(t)\in[0,1/2]$. Therefore the Bell violation
is given by
\begin{align}
B=2\sqrt{2}\sqrt{[0.9-f(t)]^2+[0.1-f(t)]^2},
\end{align}
The corresponding Bloch representation is
\begin{align}
c_1=0.8-1.6f(t),\ c_2=0.8,\ c_3=2f(t)-1,
\end{align}
Then the geometric discord can be cast as
\begin{align}
D_G=\left\{\begin{array}{cc}
\frac{(0.8-1.6f)^2+0.8^2}{4},    &  0\leq f\leq0.1 \\
\frac{(0.8-1.6f)^2+(1-2f)^2}{4}, &  0.1<f\leq0.5
\end{array}\right.
\end{align}

Fig. \ref{non-back-action} displays that Bell violation $B-2$ periodically vanishes and revives,
with its revival amplitude unchanged. This behavior is in sharp contrast with the situation in the non-Markovian
case, where the revival amplitude of Bell violation is damping gradually until the complete death.
The maximum values of $B$ and $D_G$ are attained at $t_n=n\pi/2g$ ($n=1,2,...$), while $D_G$ only vanishes at
$t_n=(2n-1)\pi/(4g)$ like its original version $D$ showed in Ref. \cite{Franco10}. It is worth noting that
the evolution of the relevant quantifiers is of great dependence on the initial state in this environment.
For instance, if all the initial eigenvalues $\lambda_\beta^s(0)$ ($\beta=\Psi,\Phi$; $s=\pm$) are less than $1/2$,
then the revival of Bell violation will never occur.
\section{Conclusion and Open questions}
First, We discuss the relationship between the violation of CHSH-Bell inequality and geometric measure of
quantum discord for Bell diagonal states. Exact lower and upper bounds of Bell violation
versus geometric discord are derived and the respective states which suffice those bounds are also characterized:
the lower and upper bounds are attained for Werner states and rank 2 Bell diagonal states respectively.
Second, the dynamical behavior of these two quantifiers is analyzed in the presence of decoherence,
including Markovian, non-Markovian, and non-back-action quantum environments, especially for extended Werner states.
Under Markovian noise (phase damping channel), sudden death of Bell violation occurs concerning with
the purity parameter $r$, but the revival never appears. Moreover, it is interesting to find that the relation
$B=2\sqrt{r^2+2D_G}$ always holds for both initial and decohered EWL states (comparing with the well-known
formula $B=2\sqrt{1+2D_G}$ for pure states). In the non-Markovian case, Bell violation periodically vanishes
and revives according to the function $P_t$ with its revival amplitude damping, while geometric discord
only vanishes at some discrete points. More importantly, we have clearly displayed that $B-2$ and $D_G$
simultaneously reach zero according to the zero points of $P_t$ and achieve extreme values at the maximum points of
$P_t$. Note that the sudden death and revival of Bell violation is closely related to the memory effect of
non-Markovian environment (the value of $\lambda/\Gamma$). Finally, we consider the time evolution of
$B-2$ and $D_G$ driven by a system-unaffected environment. Our result shows that Bell violation $B-2$ also
periodically vanishes and revives, however, with its revival amplitude unchanged, which is in sharp contrast
with the situation in the non-Markovian case. It is worth emphasizing that the maximum values of $B$ and $D_G$
are attained at the same time, and the evolution of these two quantifiers is of great dependence
on the initial state in this environment.

Yet there are still some open theoretical problems for further research. One is the question that whether
there exists a strict lower or upper bound of Bell violation versus geometric discord for more general states.
we conjecture that $B\leq2\sqrt{1+2D_G}$ holds for almost all states, but the proof may need more technical tricks.
In addition, we are seeking to investigate the relation between Bell violations (non-locality) and other correlation
quantities, for example, randomness, concerning that the Bell inequality violation can be used to certify
the presence of genuine randomness \cite{Pironio10,Acin11}.
\section*{Acknowledgements}
The authors would like to acknowledge the anonymous reviewer for his/her valuable comments and suggestions to improve the quality of the paper.
This work was supported by the National Basic Research Program of China (Grants No. 2011CBA00200 and No. 2011CB921200),
National Natural Science Foundation of China (Grant No. 60921091), the National High Technology Research and Development
Program of China (863 Program) (Grant No. 2009AA01A349) and China Postdoctoral Science Foundation (Grant No. 20100480695).
\section{Appendix}
\subsection{B and $D_{G}$ for X states}
In this appendix, we analytically present the maximum violation of Bell-CHSH inequality and geometric measure
of discord for a more general class of two-qubit states, X-structured states, which are represented in the orthonormal basis
$\{|00\rangle,|01\rangle,|10\rangle,|11\rangle \}$
\begin{equation}
\label{X-state}
\rho_{X}=
\left(\begin{array}{cccc}
\rho_{11} & 0 & 0 & \rho_{14} \\
0 & \rho_{22} & \rho_{23} & 0 \\
0 & \rho_{32} & \rho_{33} & 0 \\
\rho_{41} & 0 & 0 & \rho_{44}
\end{array}\right),
\end{equation}
where we can assume $\rho_{23}=\rho_{32}$ and $\rho_{14}=\rho_{41}$. For computational simplicity, it is
helpful to rewrite the state (\ref{X-state}) in Bloch decomposition
\begin{align}
\label{Bloch}
\rho=&\frac{1}{4}(I\otimes I+m\cdot\sigma_3\otimes I+I\otimes n\cdot\sigma_3\nonumber\\
&+\sum_{i=1}^{3}c_{i}\sigma_i\otimes\sigma_i),
\end{align}
with
\begin{align}
& c_1=2\rho_{14}+2\rho_{23},\nonumber\\
& c_2=-2\rho_{14}+2\rho_{23},\nonumber\\
& c_3=\rho_{11}-\rho_{22}-\rho_{33}+\rho_{44},\nonumber\\
& m=\rho_{11}+\rho_{22}-\rho_{33}-\rho_{44},\nonumber\\
& n=\rho_{11}-\rho_{22}+\rho_{33}-\rho_{44}.
\end{align}
According to the Horodecki criterion \cite{Horodecki95}, $B=2\sqrt{\max_{i<j}(u_i+u_j)}$ with $i,j=1,2,3$.
The three eigenvalues $u_i$ of $U=T^{T}T$ are
\begin{align}
& u_1=4(|\rho_{14}|+|\rho_{23}|)^2,\nonumber\\
& u_2=4(|\rho_{14}|-|\rho_{23}|)^2,\nonumber\\
& u_3=(\rho_{11}-\rho_{22}-\rho_{33}+\rho_{44})^2,
\end{align}
It is easy to see that $u_1$ is always larger than $u_2$, and thus the maximum violation for X states is
\begin{align}
B=\max(B_1,B_2),\quad B_1=2\sqrt{u_1+u_2},\quad B_1=2\sqrt{u_1+u_3},
\end{align}
Meanwhile, geometric measure of discord can be evaluated as
\begin{align}
D_{G}=\frac{1}{4}(c_1^2+c_2^2+c_3^3+m^2-\max\{c_1^2,c_2^2,c_3^2+m^2\}).
\end{align}
\subsection{Markovian noise}\label{Markovian}
In the standard product basis $\{|11\rangle,|10\rangle,|01\rangle,|00\rangle\}$, we can rewrite EWL states in matrix form
\begin{align}
\rho^\Phi(0) = \left(\begin{array}{cccc}
\frac{1-r}{4} & 0 & 0 & 0 \\
0 & \frac{1-r}{4}+\beta^2 r & \alpha\beta r & 0 \\
0 & \alpha\beta r & \frac{1-r}{4}+\alpha^2 r & 0 \\
0 & 0 & 0 & \frac{1-r}{4}
\end{array}\right),
\end{align}
\begin{align}
\rho^\Psi(0) = \left(\begin{array}{cccc}
\frac{1-r}{4}+\beta^2 r & 0 & 0 & \alpha\beta r \\
0 & \frac{1-r}{4} & 0 & 0 \\
0 & 0 & \frac{1-r}{4} & 0 \\
\alpha\beta r & 0 & 0 & \frac{1-r}{4}+\alpha^2 r
\end{array}\right),
\end{align}
Note that EWL states are a subset of X states, and thus they can also be represented in the form of Eq. (\ref{Bloch}).
For $\rho^\Phi(0)$, these parameters are given as follows
\begin{align}
c_1=2\alpha\beta r,\
c_2=2\alpha\beta r,\
c_3=-r,\
m=-n=(\beta^2-\alpha^2)r.
\end{align}
While for $\rho^\Psi(0)$, we have
\begin{align}
c_1=2\alpha\beta r,\
c_2=-2\alpha\beta r,\
c_3=r,\
m=n=(\beta^2-\alpha^2)r.
\end{align}
Keeping $\alpha^2+\beta^2=1$ in mind, it is straightforward to obtain the same $B$ (and $D_G$) for both $\rho^\Phi(0)$
and $\rho^\Psi(0)$
\begin{align}
B=2\sqrt{r^2+4\alpha^2\beta^2 r^2},\ D_G=2\alpha^2\beta^2 r^2.
\end{align}
As is the case for phase damping, EWL states evolve as
\begin{align}
\rho^\Phi(p) = \left(\begin{array}{cccc}
\frac{1-r}{4} & 0 & 0 & 0 \\
0 & \frac{1-r}{4}+\beta^2 r & (1-p)\alpha\beta r & 0 \\
0 & (1-p)\alpha\beta r & \frac{1-r}{4}+\alpha^2 r & 0 \\
0 & 0 & 0 & \frac{1-r}{4}
\end{array}\right),
\end{align}
\begin{align}
\rho^\Psi(p) = \left(\begin{array}{cccc}
\frac{1-r}{4}+\beta^2 r & 0 & 0 & (1-p)\alpha\beta r \\
0 & \frac{1-r}{4} & 0 & 0 \\
0 & 0 & \frac{1-r}{4} & 0 \\
(1-p)\alpha\beta r & 0 & 0 & \frac{1-r}{4}+\alpha^2 r
\end{array}\right),
\end{align}
Through Eq. (\ref{Kraus}) and (\ref{phase-damping}), we can find the only change is that $c_1$ and $c_2$
become $c'_1=(1-p)c_1$ and $c'_2=(1-p)c_2$ respectively for both $\rho^\Phi(p)$ and $\rho^\Psi(p)$, and
other parameters $c_3,m,n$ remain unchanged. Hence, $B$ (and $D_G$) are still the same with respect
to $\rho^\Phi(p)$ and $\rho^\Psi(p)$ (note that $0\leq p\leq1$)
\begin{align}
B=2\sqrt{r^2+4(1-p)^2\alpha^2\beta^2 r^2},\ D_G=2(1-p)^2\alpha^2\beta^2 r^2.
\end{align}
\subsection{non-Markovian noise}\label{non-Markovian}
Following the non-Markovian model described in Ref. \cite{Bellomo07}, we can easily obtain the exact evolution
of EWL states. For $\rho_\Phi$, the decohered nonvanishing density matrix elements are
\begin{gather}
\rho^\Phi_{11}(t)=\frac{1-r}{4}P_t^2,\nonumber\\
\rho^\Phi_{22}(t)=(\frac{1-r}{4}+\beta^2r)P_t+\frac{1-r}{4}P_t(1-P_t),\nonumber\\
\rho^\Phi_{33}(t)=(\frac{1-r}{4}+\alpha^2r)P_t+\frac{1-r}{4}P_t(1-P_t),\nonumber\\
\rho^\Phi_{44}(t)=1-P_t+\frac{1-r}{4}P_t^2,\nonumber\\
\rho^\Phi_{23}(t)=\rho^\Phi_{32}(t)=\alpha\beta\gamma P_t,
\end{gather}
The corresponding Bloch representation is also of the form Eq. (\ref{Bloch})
\begin{align}
& c_1(t)=c_2(t)=2\alpha\beta rP_t,\nonumber\\
& c_3(t)=1-2P_t+(1-r)P^2_t,\nonumber\\
& m(t)=(\beta^2-\alpha^2)rP_t-(1-P_t),\nonumber\\
& n(t)=(\alpha^2-\beta^2)rP_t-(1-P_t),
\end{align}
Meanwhile, for $\rho_\Psi$, we obtain
\begin{gather}
\rho^\Phi_{11}(t)=(\frac{1-r}{4}+\beta^2r)P_t^2,\nonumber\\
\rho^\Phi_{22}(t)=\rho^\Phi_{33}(t)=\frac{1-r}{4}P_t+(\frac{1-r}{4}+\beta^2r)P_t(1-P_t),\nonumber\\
\rho^\Phi_{44}(t)=1+(\frac{1-r}{4}+\beta^2r)P_t^2-(1-r+2\beta^2r)P_t,\nonumber\\
\rho^\Phi_{14}(t)=\rho^\Phi_{41}(t)=\alpha\beta rP_t,
\end{gather}
The corresponding Bloch decomposition is given by
\begin{align}
& c_1(t)=-c_2(t)=2\alpha\beta rP_t,\nonumber\\
& c_3(t)=1+(\frac{1-r}{4}+\beta^2r)4P_t(1-P_t)-(1-r)P_t,\nonumber\\
& m(t)=n(t)=(\beta^2-\alpha^2)rP_t-(1-P_t),
\end{align}
Exploiting the above formulas, we can conveniently carry out any dynamical simulation with respect to
all these parameters.



\begin{thebibliography}{99}
\bibitem{Bell64} J. S. Bell, Physics (Long Island City, N.Y.) \textbf{1}, 195 (1964).
\bibitem{CHSH69} J. F. Clauser, M. A. Horne, A. Shimony, and R. A. Holt, Phys. Rev. Lett.
\textbf{23}, 880 (1969).
\bibitem{Pironio10} S. Pironio, et al., Nature \textbf{464}, 1021 (2010).
\bibitem{Barrett05} J. Barrett, A. Ac\'{\i}n, N. Gisin, Phys. Rev. Lett. \textbf{95}, 010503 (2005).
\bibitem{Acin07} A. Ac\'{\i}n, N. Brunner, N. Gisin, S. Massar, S. Pironio, V. Scarani, Phys. Rev. Lett.
\textbf{98}, 230501 (2007).
\bibitem{Pironio09} S. Pironio, A. Ac\'{\i}n, N. Brunner, N. Gisin, S. Massar, V. Scarani, New J. Phys.
\textbf{11}, 045021 (2009).
\bibitem{Verstraete02} F. Verstraete and M. M. Wolf, Phys. Rev. Lett \textbf{89}, 170401 (2002).
\bibitem{Derkacz} {\L}. Derkacz and L. Jak\'{o}bczyk, Phys. Lett. A \textbf{328}, 26 (2004);
Phys. Rev. A \textbf{72}, 042321 (2005).
\bibitem{Mazzola10a} L. Mazzola, B. Bellomo, R. Lo Franco, and G. Compagno, Phys. Rev. A \textbf{81}, 052116 (2010).
\bibitem{Gisin91} N. Gisin, Phys. Lett. A \textbf{154}, 201 (1991).
\bibitem{Werner89} R. F. Werner, Phys. Rev. A \textbf{40}, 4277 (1989).
\bibitem{Munro01} W. J. Munro, K. Nemoto and A. G. White, J. Mod. Opt. \textbf{48}, 1239 (2001).
\bibitem{Ghosh01} S. Ghosh, G. Kar, A. Sen(De), and U. Sen, Phys. Rev. A \textbf{64}, 044301 (2001).
\bibitem{Miranowicz04} A. Miranowicz and A. Grudka, J. Opt. B 6, 542 (2004).
\bibitem{Ollivier01} H. Ollivier and W. H. Zurek, Phys. Rev. Lett. \textbf{88}, 017901 (2001).
\bibitem{Henderson01} L. Henderson and V. Vedral, J. Phys. A \textbf{34}, 6899 (2001).
\bibitem{Discord1}
S. Luo, Phys. Rev. A \textbf{77}, 022301 (2008); K. Modi, T. Paterek, W. Son, V. Vedral, and M. Williamson,
Phys. Rev. Lett. \textbf{104}, 080501 (2010); M. D. Lang and C. M. Caves, Phys. Rev. Lett. \textbf{105}, 150501 (2010);
S. Luo and S. Fu, Phys. Rev. A \textbf{82}, 034302 (2010); A. Ferraro, L. Aolita, D. Cavalcanti, F. M. Cucchietti,
and A. Ac\'{\i}n, Phys. Rev. A \textbf{81}, 052318 (2010).
\bibitem{Discord2} J.-S. Xu, C.-F. Li, C.-J. Zhang, X.-Y. Xu, Y.-S. Zhang, and G.-C. Guo, Phys. Rev. A \textbf{82}, 042328 (2010);
J.-S. Xu et al., Nat. Commun. \textbf{1}, 7 (2010).
\bibitem{Discord3} A. Datta, A. Shaji, and C. M. Caves, Phys. Rev. Lett. \textbf{100},
050502 (2008); M. Piani, P. Horodecki, and R. Horodecki, Phys. Rev. Lett. \textbf{100},
090502 (2008); B. P. Lanyon, M. Barbieri, M. P. Almeida, and A. G. White, Phys. Rev. Lett. \textbf{101},
200501 (2008).
\bibitem{Luo08} S. Luo, Phys. Rev. A \textbf{77}, 042303 (2008).
\bibitem{X-state} M. Ali, A. R. P. Rau, and G.Alber, Phys. Rev. A \textbf{81}, 042105 (2010);
Q. Chen, C. J. Zhang, S. X. Yu, X. X. Yi, and C. H. Oh, Phys. Rev. A \textbf{84}, 042313 (2011).
\bibitem{Girolami11} D. Girolami and G. Adesso, Phys. Rev. A \textbf{83}, 052108 (2011).
\bibitem{Dakic10} B. Daki\'{c}, V. Vedral, and C. Brukner, Phys. Rev. Lett.
\textbf{105}, 190502 (2010);
\bibitem{Batle11a} J. Batle, A. Plastino, A. R. Plastino, and M. Casas, J. Phys. A: Math. Theor. \textbf{44}, 505304 (2011).
\bibitem{Batle11b} J. Batle, M. Casas, J. Phys. A: Math. Theor. \textbf{44}, 445304 (2011).
\bibitem{Horodecki95} R. Horodecki, P. Horodecki and M. Horodecki, Phys. Lett. A \textbf{200}, 340 (1995);
R. Horodecki, Phys. Lett. A \textbf{210}, 223 (1996).
\bibitem{Bennett96} C. H. Bennett, G. Brassard, S. Popescu, B. Schumacher, J. A. Smolin, and W. K. Wootters,
Phys. Rev. Lett. \textbf{76}, 722 (1996).
\bibitem{Horodecki96} R. Horodecki and M. Horodecki, Phys. Rev. A \textbf{54}, 1838 (1996).
\bibitem{Spengler11} C. Spengler, M. Huber and B. C. Hiesmayr, J. Phys. A: Math. Theor, \textbf{44}, 065304 (2011).
\bibitem{Mazzola10b} L. Mazzola, J. Piilo, and S. Maniscalco, Phys. Rev. Lett. \textbf{104}, 200401 (2010).
\bibitem{Maziero09} J. Maziero, L. C. Celeri, R. M. Serra, and V. Vedral, Phys. Rev. A \textbf{80}, 044102 (2009).
\bibitem{Werlang09} T. Werlang, S. Souza, F. F. Fanchini, and C. J. Villas Boas, Phys. Rev. A \textbf{80}, 024103 (2009).
\bibitem{Fanchini09} F. F. Fanchini, T. Werlang, C. A. Brasil, L. G. E. Arruda, and A. O. Caldeira,
Phys. Rev. A \textbf{80}, 024103 (2009).
\bibitem{Wang10} B. Wang, Z.-Y. Xu, Z.-Q. Chen, and M. Feng, Phys. Rev. A \textbf{81}, 014101 (2010).
\bibitem{Lu10} X. M. Lu, Z. J. Xi, Z. Sun, and X. Wang, Quant. Inform. Comput. \textbf{10}, 0994 (2010).
\bibitem{Xu10} Z. Y. Xu, W. L. Yang, X. Xiao, and M. Feng, J. Phys. A: Math. Theor. \textbf{44}, 395304 (2011).
\bibitem{Bell-decoherence} A. Miranowicz, Phys. Lett. A \textbf{327}, 272 (2004);
L. Jak\'{o}bczyk and A. Jamr\'{o}z, Phys. Lett. A \textbf{318}, 318 (2003);
A. G. Kofman and A. N. Korotkov, Phys. Rev. A \textbf{77}, 052329 (2008);
Y. Yeo, J. H. An, and C. H. Oh, Phys. Rev. A \textbf{82}, 032340 (2010);
B. Bellomo et al., Int. J. Quantum Inf. \textbf{9}, 63 (2011);
L. Mazzola, Phys. Scr. \textbf{T140}, 014055 (2010).
\bibitem{Kraus} K. Kraus, States, Effect, and Operations: Fundamental Notions in Quantum Theory (Springer-Verlag, Berlin, 1983).
\bibitem{Nielsen} M. A. Nielsen and I. L. Chuang, Quantum Computation and Quantum Information
(Cambridge University Press, Cambridge, 2000).
\bibitem{Wootters} S. Hill and W. K. Wootters, Phys. Rev. Lett. \textbf{78}, 5022 (1997) ;
W. K. Wootters, Phys. Rev. Lett. \textbf{80}, 2245 (1998).
\bibitem{Bellomo08} B. Bellomo, R. Lo Franco, and G. Compagno, Phys. Rev. A \textbf{78}, 062309 (2008).
\bibitem{Jaeger08} G. Jaeger and K. Ann, Phys. Lett. A \textbf{372}, 2212 (2008).
\bibitem{Yu} T. Yu and J. H. Eberly, Phys. Rev. Lett. \textbf{93}, 140404 (2004); Phys. Rev. Lett. \textbf{97}, 140403 (2006);
Science \textbf{323}, 598 (2009).
\bibitem{Bellomo07} B. Bellomo, R. Lo Franco, and G. Compagno, Phys. Rev. Lett. \textbf{99}, 160502 (2007).
\bibitem{Franco10} R. Lo Franco, B. Bellomo, E. Andersson, and G. Compagno, Phys. Rev. A \textbf{85}, 032318 (2012).
\bibitem{Zyczkowski01} K. Zyczkowski, P. Horodecki, M. Horodecki, and R. Horodecki, Phys. Rev. A \textbf{65}, 012101 (2001).
\bibitem{Alicki} R. Alicki and K. Lendi, Quantum Dynamical Semigroups and Applications,
Lect. Notes Phys. 717 (Springer, Berlin Heidelberg, 2007).
\bibitem{Acin11} A. Acin, S. Massar, and S. Pironio, Phys. Rev. Lett. 108, 100402 (2012).
\end{thebibliography}
\end{document}